\documentclass[10pt,conference]{IEEEtran}
\IEEEoverridecommandlockouts

\usepackage{cite}
\usepackage{amsmath,amssymb,amsfonts}
\usepackage{algorithmic}
\usepackage{graphicx}
\usepackage{textcomp}
\usepackage{xcolor}
\usepackage{booktabs}
\usepackage{pifont}
\usepackage{tcolorbox}
\usepackage{tabularx}
\usepackage{subcaption}
\tcbuselibrary{skins}
\def\BibTeX{{\rm B\kern-.05em{\sc i\kern-.025em b}\kern-.08em
    T\kern-.1667em\lower.7ex\hbox{E}\kern-.125emX}}

\newcommand{\cmark}{\ding{51}}
\newcommand{\xmark}{\ding{55}}
\newcounter{finding}

\newcommand{\finding}[1]{%
\refstepcounter{finding}%
\begin{tcolorbox}[
  colback=gray!10,
  colframe=gray!50,
  boxrule=0.4pt,
  arc=1pt,
  left=6pt,right=6pt,top=4pt,bottom=4pt,
  fontupper=\small,
  before skip=8pt,after skip=8pt,
  enhanced
]
\textbf{Finding \thefinding:} #1
\end{tcolorbox}
}
    
\begin{document}

\title{On the risk of coding before testing: An empirical study on LLM-based test generation workflow}

\author{\IEEEauthorblockN{Michael Konstantinou}
\IEEEauthorblockA{
\textit{SnT, University of Luxembourg}\\
Luxembourg \\
michael.konstantinou@uni.lu}
\and
\IEEEauthorblockN{Florian Tambon}
\IEEEauthorblockA{
\textit{SnT, University of Luxembourg}\\
Luxembourg \\
florian.tambon@uni.lu}
\and
\IEEEauthorblockN{Mike Papadakis}
\IEEEauthorblockA{
\textit{SnT, University of Luxembourg}\\
Luxembourg \\
michail.papadakis@uni.lu}
}

\maketitle

\begin{abstract}
Large Language Models (LLMs) are increasingly used in software engineering workflows to automatically generate both source code and corresponding test suites. This dual capability has enabled emerging development paradigms, including test-first and agentic workflows, where a single model is responsible for producing and validating implementations. However, these approaches implicitly assume that generated tests act as independent and reliable oracles—a fundamental requirement for effective software testing. In this paper, we challenge this assumption and investigate whether LLM-generated code biases the generation of subsequent tests. We introduce and empirically study the phenomenon of error propagation, where faults present in generated code are systematically replicated in the associated test artifacts. This leads to cases where incorrect implementations and tests are mutually consistent, thereby masking defects rather than revealing them. We evaluate this effect across a range of programming tasks and agentic workflows, analyzing the consistency between generated code and test assertions, with particular focus on scenarios of aligned failures. Our study examines (i) whether erroneous code artifacts bias test generation, (ii) whether such bias persists under different prompting strategies, including chain-of-thought reasoning, and (iii) how errors propagate across multi-step workflows in which intermediate outputs are reused as context. The results show that error propagation is both prevalent and impactful: generating tests after faulty code significantly reduces fault detection effectiveness compared to generating tests independently (14\% vs. 25\%). These findings highlight a fundamental limitation of current workflows, where lack of independence between generated artifacts undermines the reliability of automated testing. Furthermore, our results expose a previously underexplored threat to validity in empirical studies that rely on coupled generation pipelines.
\end{abstract}


\section{Introduction}
Large Language Models (LLMs) are increasingly integrated into software engineering workflows, where they are used not only to generate production code but also to synthesize accompanying test suites. This dual capability has fueled optimistic visions of highly automated development pipelines, in which both implementation and validation are delegated to the same generative system. In particular, recent toolchains promote test-first or co-evolutionary workflows, where LLMs generate candidate tests and subsequently produce code that satisfies them, potentially accelerating development while reducing human effort.
However, this paradigm implicitly assumes that LLM-generated tests provide an independent and reliable oracle for assessing correctness. In traditional software testing, effectiveness critically depends on the independence between the system under test and the test oracle \cite{0020331,0017273}. When both artifacts are produced by the same underlying model, this assumption may no longer hold.

In this paper, we investigate this fundamental assumption by examining whether the generation of faulty code by an LLM biases the generation of tests. When tasked with producing both artifacts for the same problem, an LLM is likely to rely on similar internal representations, assumptions, and reasoning trajectories. This shared generative process increases the likelihood that errors are not independent but systematically replicated, leading to tests that encode and validate the same incorrect behavior exhibited by the generated code.

Moreover, this issue is exacerbated by the autoregressive nature of LLMs. These models generate outputs token by token, conditioning each step on previously generated content \cite{sengupta2026contextdependencereliabilityautoregressive,cheng2026contextualdragerrorscontext}. As a result, when an LLM produces code containing errors, this faulty output becomes part of the context that shapes subsequent generations. Consequently, follow-up activities—such as test generation—can be implicitly biased by earlier mistakes, increasing the risk of propagating erroneous assumptions across artifacts.

This implies that errors introduced during code synthesis—such as incorrect interpretations of specifications, boundary conditions, or edge cases—are not independent from subsequent testing activities. Instead, they are often replicated during test generation, including in the construction of test oracles (e.g., assertions). This behavior is particularly critical in agentic workflows, especially in ``vibecoding'' scenarios where autonomous agents iteratively solve tasks with minimal human intervention. Furthermore, it raises concerns for empirical studies that do not isolate code generation from test generation, as such coupling may systematically bias evaluation results.

We refer to this phenomenon as \textit{error propagation}, whereby implementation faults are mirrored by corresponding faults in generated tests. In such cases, the tests validate incorrect behavior rather than expose it, creating a false sense of correctness. Importantly, this is not merely a consequence of poor prompt design or training data limitations, but a structural effect of autoregressive generation, where similar probabilistic biases and reasoning patterns are reused across related outputs.

Recent research has shown that incorporating erroneous artifacts, such as faulty code, into prompts can degrade LLM performance. In the context of test oracle generation, studies have demonstrated that the inclusion of incorrect or misleading code during prompting, can negatively affect the correctness of generated assertions and reduce their effectiveness as validation mechanisms \cite{huang2024measuring, mizrahi2024state,konstantinou2024llmsgeneratetestoracles}. Building on these findings, we argue that the impact of erroneous artifacts extends beyond individual tasks to multi-step workflows, which more closely reflect current practice in modern agentic systems and ``vibecoding.'' In such settings, intermediate outputs—potentially containing errors—are explicitly reused as part of the context in subsequent steps, forming iterative feedback loops that shape future generations. 

Moreover, most contemporary agentic workflows rely on prompt decomposition or planning strategies, where complex tasks are broken down into sequences of sub-tasks processed step by step. While such structured approaches generally improve performance compared to single-shot prompting, they introduce strong interdependencies between steps, since each sub-task conditions on artifacts produced earlier in the workflow. Consequently, the assumption of independence between reasoning steps no longer holds. Errors introduced in earlier stages may bias or constrain subsequent generations, leading to cascading effects across the workflow.

Given that workflow design is a critical factor influencing the performance of LLM-based systems, we conduct an empirical study of agentic code-and-test generation workflows across a diverse set of programming tasks. Specifically, we analyze the consistency between generated implementations and their corresponding test assertions, with particular emphasis on scenarios in which both artifacts are incorrect yet mutually consistent. Such cases are particularly problematic, as they may falsely indicate correctness despite underlying faults.

To investigate this phenomenon, we first examine whether the inclusion of erroneous code artifacts biases the correctness of subsequently generated tests and their associated oracles. We then evaluate whether this bias persists under different prompting strategies, including structured reasoning approaches such as chain-of-thought prompting, which explicitly guide the model through intermediate steps.

After establishing the presence of this effect, we further assess the extent to which it propagates across multi-step agentic workflows. In particular, we study whether errors introduced during earlier stages—such as incorrect code generation—carry over to later stages, such as test generation, thereby inducing systematic biases. This allows us to characterize how error-prone intermediate artifacts influence downstream components within realistic, step-wise LLM-driven development pipelines.

Our results show that such aligned failures occur frequently, significantly reducing the effectiveness of testing as a validation mechanism. These failures occur even when prompting techniques are applied, in a sense chain-of-thought prompting appear to have no effect. In particular, generating tests before code generation achieves a fault detection rate of 25\%, significantly surpassing the fault detection rate of approximately 14\% observed when tests are generated after erroneous code.

These findings have important implications for the design of AI-assisted software engineering tools. They suggest that generating both code and tests using the same model does not guarantee meaningful verification, and may instead introduce a new class of undetected faults. This challenges common assumptions about the benefits of LLM-driven testing and highlights the need for approaches that introduce independence, diversity, or external grounding in the validation process.

Perhaps more importantly, our results reveal a fundamental threat to validity in LLM-based evaluation protocols. Specifically, the influence of earlier workflow steps on subsequent ones can artificially distort performance measurements, affecting both proposed and baseline approaches. While prior work has highlighted the negative effects of erroneous prompts, no previous study has systematically investigated the impact of error propagation across workflow steps, which lies at the core of modern agentic systems.

\section{Background}

\subsection{Agentic systems}

Agentic systems can be abstracted down as an LLMs which have been given the capacity to interact and act with its surrounding environments. Where traditional LLMs involves a fixed context (``prompt") on which to predict, an agent instead will allow the LLM to interact with a user by leveraging a memory (to retain contextual information), a planning capability (reasoning) and access to various tools with which it can interact with its environment. In their current forms, agentic systems are based on the ReActIng framework, that is Reasoning, Acting, Interacting \cite{PlaatDSPPB25}. In this paper, we focus mainly on the impact over the Reasoning part since we are interested in the workflow step the agent takes to solve the problem. There are several different ways of applying reasoning to an agent for test generations \cite{abs-2402-02716} and so we will focus on techniques that have been used in test-generation. While advanced agentic systems might combine several of these techniques, in our study use the techniques in isolation to assess its individual contribution to performance. The Acting and Interacting parts will be simulated via a simple process: the agent will be allowed to execute the tests generated on the code, receiving feedback and modifying its answer following established LLM prompting methodologies.

\subsection{LLM-Based test generation}

Agentic pipelines include an interaction loop where LLMs act as autonomous agents within the software environment \cite{hayet2024chatassert}. In these architectures, prompting via zero-shot, executing and iteratively repairing test code based on execution feedback or compilation errors represents the agent's 'acting' and 'interacting' phase. However, said methods quickly hit a bottleneck as they lack a proper 'reasoning' phase required to understand complex semantics before action is taken. To bridge this gap, contemporary frameworks inject explicit reasoning abstractions into the agentic context. For instance, techniques like ChatAssert \cite{HayetSd25} leverage automated code summarization to distill focal method functionalities into natural language, directly guiding the LLM’s contextual understanding during oracle generation. Methods such as Chain-of-Thought, which force the agent to explicitly map out execution paths or code structure before generating syntactically valid test assertions \cite{abs-2602-03181}. By shifting from unconstrained, trial-and-error generation to structured, execution-free semantic analysis, these reasoning mechanisms mitigate hallucinations and counteract the logical fragility of LLM code generation.

\section{Research questions}

Recent advances in LLM-based test generation have introduced a wide range of strategies that aim to improve the quality of generated test suites. Despite their differences, these approaches almost universally rely on exposing the implementation under test to the language model, either explicitly by providing the source code or implicitly. While this design choice has become common practice~\cite{konstantinou2026llmbasedtestgenerationtechniques}, its impact on fault detection effectiveness remains largely unexplored, particularly in modern LLM-assisted programming scenarios such as agentic software engineering (e.g. vibe coding). 

We therefore begin by investigating whether exposing the implementation under test actually benefits LLM-based test generation and whether the optimization strategies proposed in prior work provide additional value beyond generating tests directly from the task specification. Thus, we ask:

\textbf{RQ1 (Implementation influence): } \textit{To what extent does exposure to implementation code, either explicit or implicit, influence the ability of LLM-generated unit tests to detect faults?}

We investigate this question by checking the differences on the number of detected faults in the presence/absence of code from the test generation prompting. In a sense, we check whether test generation can actually detect the faults generated by the LLM when generating code from the same prompt (task description) if we include the generated code or not.  

The purpose of RQ1 is to check for an effect, i.e., faulty implementations bias test generation. Given that we found evidence that the inclusion of the faulty implementations during prompting biasses the test generation, negatively affects it, we move on to investigate whether the same phenomenon persists when using different prompting strategies, such as Test via Summarization, Chain-of-Thought (CoT) and Chain-of-Verification (CoVe). We therefore ask:

\textbf{RQ2 (Prompting): } \textit{To what extent does exposure to implementation code, influence the ability of LLM-generated unit tests to detect faults under different prompting strategies such as Test via Summarization, CoT and CoVe?}

We investigate this question by checking the differences on the number of detected faults in the presence/absence of code from the different prompting strategies similar to the analysis we performed in RQ1.  

While the first two research questions quantify the harmful effects of the inclusion of erroneous code in the prompt under analysis, they do not show that workflows, past tasks in particular, implicitly bias in the followup activities. Hence we ask:

\textbf{RQ3 (Workflows): } \textit{How does a typical agentic workflow compare against a workflow lacking the presence of implementation?}

We investigate this question by checking the differences on the number of detected faults when generating tests before any code generation and after. In a sense comparing the implicit bias posit by the code generation trial.

\section{Experimental setup}

\subsection{Methodology}
\label{sec:methodology}

\begin{figure*}[t]
    \centering
    \includegraphics[width=0.75\linewidth]{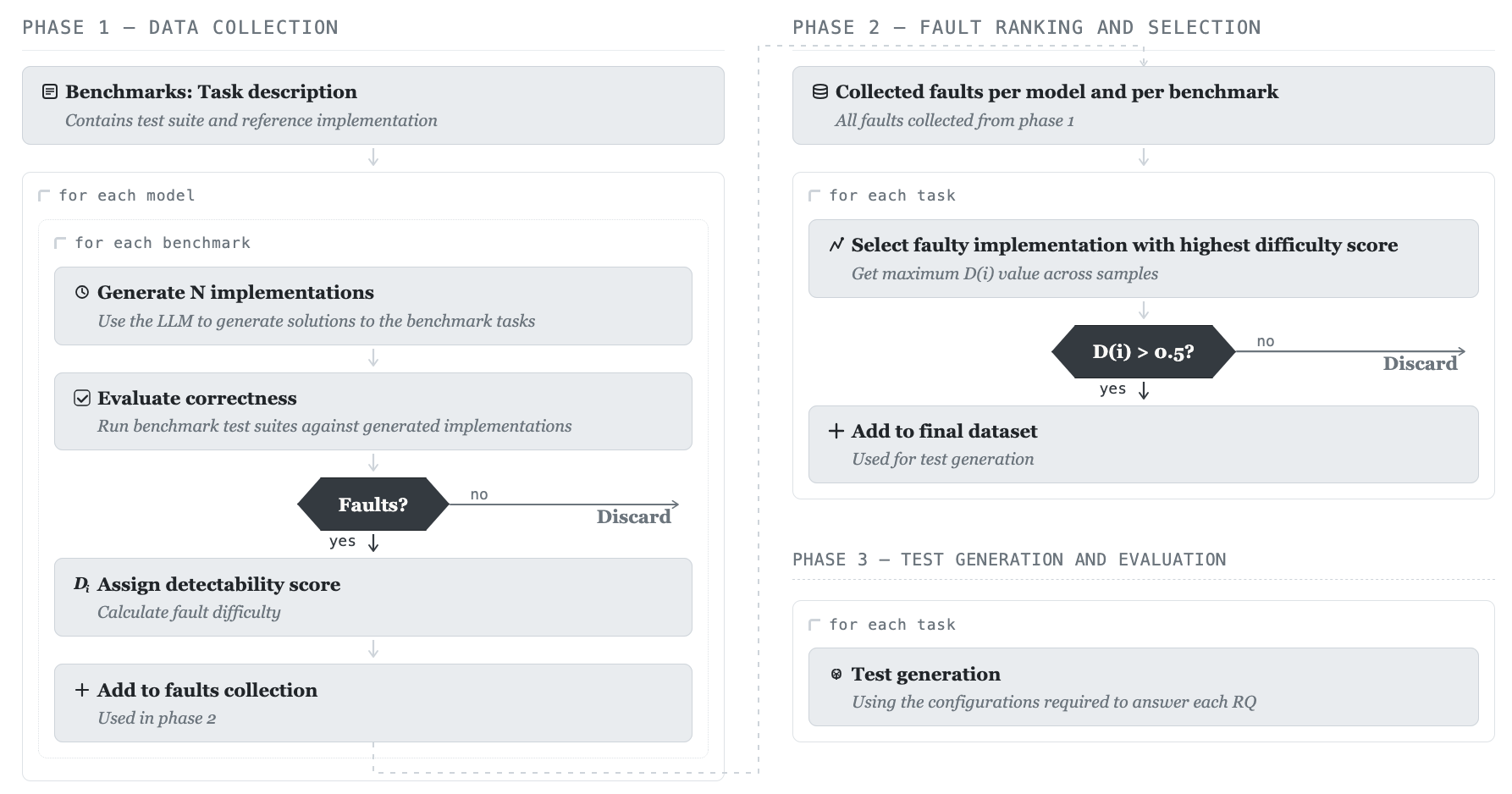}
    \vspace{-0.5em}
    \caption{\textbf{Methodology.} Faulty implementations are generated and filtered to construct the evaluation dataset. The resulting implementations are subsequently used to evaluate the effectiveness of the investigated LLM-based test generation techniques.}
    \vspace{-0.5em}
    \label{diagram:methodology}
\end{figure*}

Given a programming task, a developer can leverage an LLM or an LLM-based coding assistant to generate an implementation from a natural-language specification. Subsequently, either the developer or an automated tool may prompt the same model to generate unit tests with the goal of validating the implementation and uncovering potential faults.

Numerous approaches have been proposed, both in academia and industry, for generating tests using LLMs. Despite their differences, these approaches typically rely, either explicitly or implicitly, on access to the implementation under test. The generated code is therefore often used as the primary source of information for guiding test generation.

In this work, we investigate whether access to the implementation is necessary for effective test generation. Specifically, we compare existing code-aware test generation workflows against a simple baseline that relies exclusively on the original task description, without any access to the generated implementation. By doing so, we assess the extent to which the natural-language specification alone contains sufficient information to guide fault-revealing test generation.

To conduct this study, we perform the following steps. First, for each programming task, we use an LLM to generate one or more candidate implementations. Second, we identify and retain only faulty implementations, thereby constructing a benchmark of incorrect solutions. Finally, we apply a range of test generation workflows to each faulty implementation and evaluate their effectiveness in detecting faults. This experimental design enables a systematic comparison between implementation-aware and specification-driven test generation approaches, providing insights into the relative value of code and task descriptions as sources of testing information.

\textbf{Data Collection:} To construct our dataset of faulty implementations, we employ three well-established programming benchmarks. Each benchmark provides a task description, a reference implementation, and a test suite for validating generated solutions. For every task, we prompt each model to generate an implementation based solely on the task description. To obtain a diverse set of candidate implementations while mitigating sampling bias, we generate ten solutions per task using a temperature of 0.8, which encourages variation in the models' outputs.

Each generated implementation is executed against the benchmark's reference test suite. Since our study focuses on the effectiveness of generated tests in detecting behavioral faults, we retain only implementations that produce incorrect outputs for at least one test case. Any implementations that fail due to runtime errors, exceptions, or other execution failures are excluded.

\textbf{Faults ranking and selection:} The previous phase produces a set of candidate faulty implementations for each programming task. However, multiple faulty implementations may be generated for the same task, and some may contain faults that are trivially detectable. 

To obtain a representative and challenging evaluation dataset, we apply the following filtering criteria to the generated implementations.

\begin{enumerate}
\item \textit{Fault ranking.} Each implementation is assigned a difficulty score based on the proportion of reference test cases that expose its fault. Specifically, we use the number of failing tests as a proxy for fault detectability: implementations that fail a large fraction of the benchmark's test suite are considered easier to detect, whereas implementations that fail only a small number of test cases are considered more difficult to expose, as they require more targeted test inputs.

\item \textit{Fault filtering.} Our objective is to evaluate test generation techniques on implementations whose faults are not trivially exposed by the benchmark's reference test suite. Therefore, we exclude implementations for which more than 50\% of the reference test cases fail. This threshold serves as a pragmatic filtering criterion to remove implementations whose faults are already detected by a large fraction of the available tests, allowing the evaluation to focus on more challenging fault scenarios.

\item \textit{Implementation selection.} For each programming task, we retain a single faulty implementation, if one exists. When multiple candidates exist, we select the implementation with the highest estimated difficulty (i.e., the one that fails the fewest reference test cases), thereby prioritizing faults that are less readily exposed by existing tests.

\end{enumerate}

\textbf{LLM-based test generation:} To evaluate the impact of implementation exposure on test generation, we require a test generation workflow that minimizes confounding factors and allows differences in performance to be attributed solely to the test strategy under investigation. We therefore adopt \textit{LLM-Plain}~\cite{konstantinou2026llmbasedtestgenerationtechniques}, a lightweight test generation approach that relies solely on the reasoning capabilities of the underlying LLM. 

Most prior techniques target the creation of the so-called regression tests \cite{huang2024measuring}, and therefore assume the behavior of the code under test to be correct, i.e., they cannot reveal faults by construction \cite{konstantinou2024llmsgeneratetestoracles}. LLM-Plain includes an iterative refinement loop, similar to those employed by modern LLM-based test generation tools and agentic coding workflows, in which the model iteratively corrects invalid outputs. In our study, we modify this refinement loop to repair only compilation errors. This ensures that generated test suites are executable, with assertions (oracles) been selected by the LLM with the intention to reveal candidate faults. Prior work has shown that LLM-Plain achieves competitive performance with state-of-the-art LLM-based test generation tools~\cite{konstantinou2026llmbasedtestgenerationtechniques}, making it an appropriate baseline for our evaluation.

\subsection{Benchmarks}

To evaluate the test generation techniques and workflows presented in this study, we employ three widely adopted code generation benchmarks. Each benchmark consists of a collection of programming tasks, comprising a natural-language problem specification, a reference implementation, and an executable test suite for validating candidate solutions. Since these benchmarks contain only correct reference implementations, we use them to construct a dataset of faulty implementations following the methodology described in Section~\ref{sec:methodology}. The accompanying reference test suites are then used both to identify faulty implementations and to evaluate the fault detection effectiveness of the test suites generated by each technique. Specifically, we use the following benchmarks:

\begin{itemize}

\item \textbf{HumanEval+:}~\cite{liu2024your}~\cite{chen2021evaluating} HumanEval+ is an extension of the widely used HumanEval benchmark that augments each programming task with a significantly more comprehensive test suite. It contains 164 Python programming problems specified in natural language, accompanied by a reference implementation and an extensive set of tests.

\item \textbf{MBPP:}~\cite{austin2021program} Mostly Basic Python Problems (MBPP) is a benchmark of 974 crowd-sourced Python programming tasks covering a broad range of introductory programming concepts.

\item \textbf{BigCodeBench:}~\cite{zhuo2025bigcodebench} BigCodeBench is a benchmark designed to evaluate LLMs on more realistic and challenging programming tasks. Compared to HumanEval+ and MBPP, it contains substantially more complex problems that often require multiple reasoning steps and the use of external libraries, providing a closer approximation to real-world software development. The benchmark consists of 1,140 programming tasks spanning 139 Python libraries.

\end{itemize}

\subsection{Models}

We employ 5 state-of-the-art language models from different providers: GPT-5-mini, GPT-4.1-mini, DeepSeek-V4-Flash, Claude Haiku 4.5, and Llama 3.3 Instruct (70B). The selected models represent a diverse range of capabilities, including both innate reasoning-enabled and non-reasoning models, as well as open-weight and commercially available models. This diversity enables the generalization assessment of the observed effects across different model families and deployment settings. Table~\ref{tab:models} lists the models we used, indicating whether each model is open-weight and whether its reasoning capabilities are supported and were used in the study.

\begin{table}[ht]
\centering
\caption{Language models used alongside their open-weight availability and their reasoning capabilities used}
\vspace{-0.5em}
\label{tab:models}
\begin{tabular}{lcc}
\toprule
\textbf{Model} & \textbf{Open-weight} & \textbf{Reasoning} \\
\midrule
GPT-5-mini               & \xmark & \cmark \\
GPT-4.1-mini             & \xmark & \xmark \\
DeepSeek-V4-Flash        & \cmark & \xmark \\
Claude Haiku 4.5         & \xmark & \cmark \\
Llama 3.3 Instruct (70B) & \cmark & \xmark \\
\bottomrule
\end{tabular}
\vspace{-0.5em}
\end{table}

\subsection{Test generation strategies}

Modern LLM-based test generation tools typically combine multiple prompting strategies, iterative refinement loops, and auxiliary analyses to improve the quality of the generated test suites. As we do not aim to evaluate the tools but rather the impact of their use within different workflows. The selected strategies reflect the current state of practice in LLM-based test generation, as each has been proposed in prior work or adopted by existing test generation tools. 

\textbf{Baseline:} As we aim to study reasoning workflow impact on agent's code-test-generation performance, we consider a simple agentic baseline \textit{without} any added reasoning capability. 
We use LLM-Plain~\cite{konstantinou2026llmbasedtestgenerationtechniques}, a simple iterative error-refinement loop to ensure test suite are executable. This ensures some agentic capability without including any added reasoning capability and enables us to isolate and evaluate the contribution of each strategy independently.

\textbf{Test via Summarization:} Test summarization is a strategy that first prompts the LLM to summarize the code under test before generating the corresponding test suite. 

The intuition is that a natural-language summary guides the model to capture the intended behavior. This strategy was introduced by ChatAssert~\cite{hayet2024chatassert} and aims at generating test assertions.

\textbf{Chain-of-Thought (CoT)~\cite{3600270.3602070}:} Before generating the final test suite, the LLM is encouraged to explicitly reason about the expected behavior of the implementation and identify relevant scenarios to test. The resulting reasoning is then used to guide the subsequent generation of unit tests. CoT has been successfully incorporated into code generation workflows~\cite{li2025structured}.

\textbf{Chain-of-Verification (CoVe)~\cite{dhuliawala2024chain}:} Unlike \textit{Chain-of-Thought}, which encourages reasoning before generating a response, \textit{Chain-of-Verification} introduces a self-verification phase after an initial response has been produced. The LLM first generates a candidate test suite and is subsequently prompted with a series of verification questions that encourage it to identify and correct potential mistakes. CoVe and similar designs were applied to improve code and test generation through several studies~\cite{konstantinou2025yate, taherkhani2026consistency, kouemo2024chain}.

\textbf{Agentic workflow:} Unlike the previous techniques, this workflow mimics an interactive session from the initial code generation request. The model first receives a code generation request and produces the (faulty) implementation. The user then asks the same model to generate unit tests. To achieve this, we preserve the entire conversation history from code generation through test generation within a single session. To ensure that all workflows are evaluated on identical faulty implementations, we control the model's initial code generation response to exactly match the faulty implementation(s) collected at an earlier stage.

\subsection{Metrics}

We evaluate the effectiveness of generated test suites using two complementary metrics. \emph{Fault triggering} measures whether a generated test exercises faulty behavior in an implementation, whereas \emph{fault detection} additionally requires that the observed failure is attributable to a correct test oracle.

\textbf{Fault Triggering:} A fault is considered \emph{triggered}  if a generated test triggers behavior that is different from that of the reference implementation, irrespective of the exact assertion statement used by the test. Strictly speaking, a fault is considered triggered if there is a behavioral difference, detected through the program outputs that are checked by the test assertions, between the faulty reference implementations, irrespective of whether the fault is detected or not. For instance, a test assertion may fail on both reference and faulty implementations but still trigger the fault if we detect that the two programs provide different outputs. This means that fault triggering is a weaker pre-condition of fault detection with any difference between the fault triggering and detected cases indicating wrongly expressed assertions.

\textbf{Fault Detection:} A fault is considered \emph{detected} if a test fails when executed against the faulty implementation while passing on the corresponding reference implementation (assumed to be correct). In other words, a fault is \emph{detected} if it is triggered \emph{and} and is correctly detected by the test. 

\textbf{Fault Detectability:} We introduce a \emph{fault detectability} metric to estimate how easily a fault is triggered by a test suite. Intuitively, a fault that is exercised by many test cases is easier to detect than one that requires highly specific inputs and output assertions. Accordingly, we estimate fault detectability using the following score: $D(i) = 1 - \frac{|F_i|}{|T|}$, where $T$ denotes the test suite and $F_i \subseteq T$ is the subset of test cases that fail for implementation $i$. Higher values of $D(i)$ correspond to implementations whose faults are exposed by fewer reference test cases and are therefore considered less readily detectable. We use this metric exclusively during dataset construction to rank faulty implementations and exclude implementations whose faults are trivially exposed, as described in Section~\ref{sec:methodology}.

\textbf{Statistical significance:} We employ the Mann-Whitney U test~\cite{mann1947test} to determine whether the results we observe are statistically significant or not. Following standard practice, we consider a result as statistical significant when the resulting $p$-value is below 0.05. 

\section{Research protocol}

To answer RQ1, we conduct three complementary experiments that evaluate the impact of code exposure on LLM-based test generation. In all experiments, we employ the LLM-Plain workflow described in Section~\ref{sec:methodology}, ensuring that the evaluated strategy is the only aspect that is controlled in the different configurations.

We first evaluate the direct impact of exposing the implementation under test during test generation. Unlike prior work~\cite{huang2024measuring}, which compares prompting strategies using a single interaction with the LLM, we perform this comparison within a realistic LLM-based test generation workflow that reflects the design of modern test generation approaches. We generate test suites using the three input configurations: (1) Task description only, (2) Task description and implementation under test, and (3) Implementation under test only. It is noted that each configuration and task is performed on a fresh trial to avoid having potential bias from previous runs. 

To answer RQ2, we investigate whether prompt engineering techniques commonly adopted by recent LLM-based test generation and code generation workflows mitigate the impact of implementation exposure. We compare the task-description-only baseline against three representative strategies: summarization-based test generation \cite{hayet2024chatassert}, Chain-of-Thought prompting \cite{3600270.3602070}, and Chain-of-Verification \cite{dhuliawala2024chain}.

 To answer RQ3, we simulate a typical vibe-coding (agentic) workflow in which a developer first prompts the LLM to generate an implementation from a task description and subsequently requests a unit test suite for the generated code. To faithfully reproduce this interaction, we preserve the conversational history throughout the workflow. Specifically, the model first receives the task description and generates an implementation. Since we are interested on what happens when the model generates faulty code, we only consider these cases, i.e., we discard the cases that the model is generating correct code and keep only those that are faulty.  We then prompt the model to generate unit tests. 
 
 This setup reflects the behavior of contemporary LLM-based coding assistants, where the model retains access to previous interactions.

We compare this workflow against the task-description workflow introduced in RQ1. In contrast to the agentic workflow, test generation is performed in a fresh interaction with the LLM that contains only the task description and no conversational context from the preceding code generation process. Consequently, the generated implementation is not exposed to the model, either explicitly or implicitly. Throughout the remainder of the paper, we refer to the former as the \emph{Agentic Workflow} and the latter as the \emph{Test-Driven Workflow}, as test generation is performed independently of the code generation.

\section{Results}

\subsection{RQ1: Implementation influence}

\begin{figure}[t]
    \centering
    \includegraphics[width=0.98\linewidth]{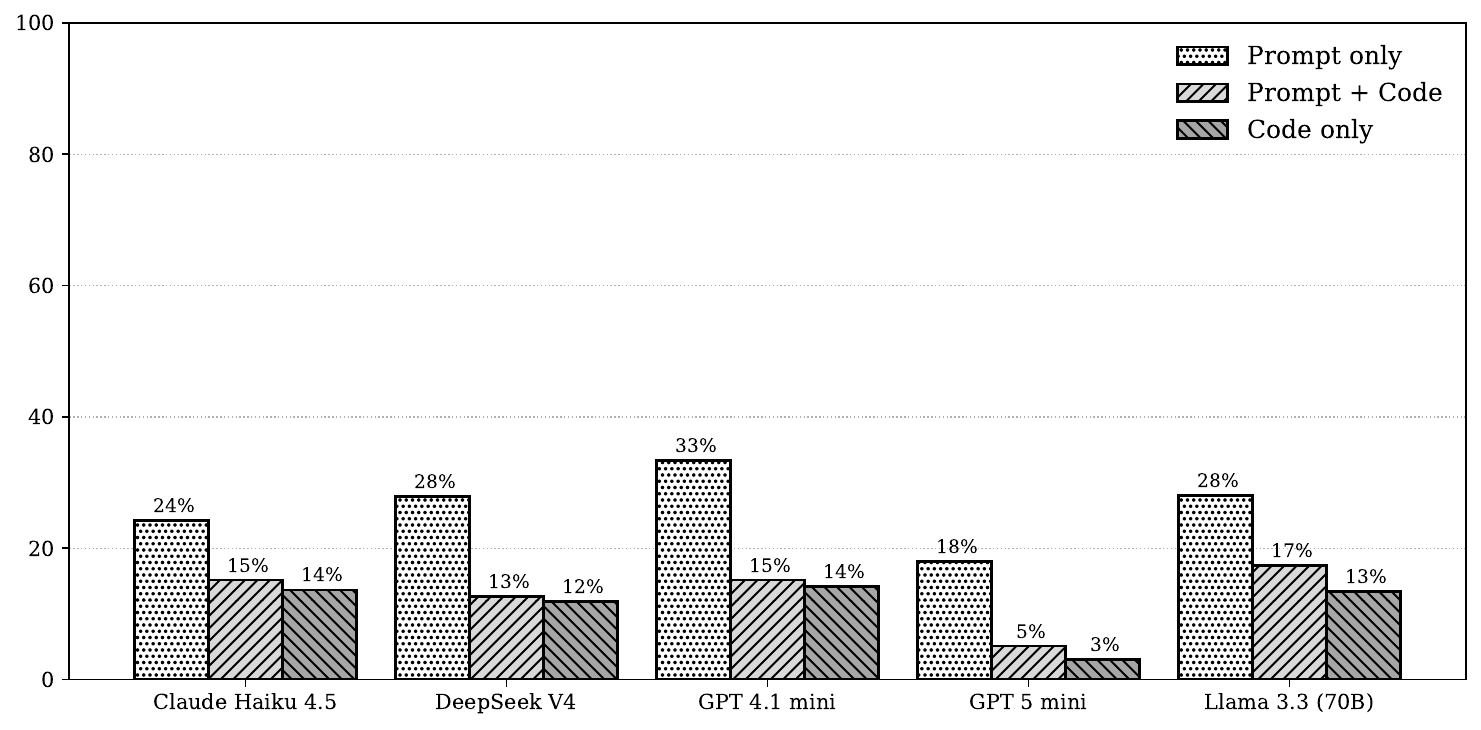}
    \caption{\textbf{RQ1: Fault Detection across all models and benchmarks.} Generating test cases from the task description alone leads to better fault detection rate.}
    \label{fig:fd_rq1_a}
    \vspace{-0.5em}
\end{figure}

Figure~\ref{fig:fd_rq1_a} presents the fault detection effectiveness when the LLM was given only the task description, the code only, or both sources of information. We observe that in all cases, the test suites that were generated directly from the task description detected more faults. Specifically, providing the implementation under test alongside the task description reduced fault detection by 9.1\%, 15.2\%, 18.2\%, 12.9\%, and 10.7\% for Claude Haiku 4.5, DeepSeek V4, GPT-4.1 mini, GPT-5 mini, and Llama 3.3 (70B), respectively. 

An interesting observation is that once the implementation under test is exposed to the model, the accompanying task description provides only a marginal improvement. On average, the difference between generating tests from the implementation alone and from both the task description and the implementation is only 1.9\%, whereas providing the task description alone yields a 13.2\% improvement over the combined configuration. This suggests that once the implementation is introduced to the model, it primarily relies on the implementation rather than the task description. 

Table~\ref{tab:rq1_stats} confirms the observations discussed above. The table summarizes the statistical significance of the pairwise comparisons. Across all evaluated models, the differences between the task-description-only configuration and the two configurations that expose the implementation under test are statistically significant. 

In contrast, no statistically significant difference is observed between generating tests from both the task description and the implementation and generating tests from the implementation alone.

\begin{table}[t]
\centering
\caption{RQ1: Statistical significance of the pairwise comparisons between the three configurations: Prompt-only (P), Prompt and Code (P+C) and Code only (C)}
\label{tab:rq1_stats}

\resizebox{\columnwidth}{!}{%
\begin{tabular}{lccc}
\toprule
\textbf{Model} &
\textbf{P vs. P+C} &
\textbf{P+C vs. C} &
\textbf{P vs. C} \\
\midrule
Claude Haiku 4.5 &
\cmark\ ($p=0.009$) &
\xmark\ ($p=0.621$) &
\cmark\ ($p=0.002$) \\

DeepSeek V4 &
\cmark\ ($p<0.001$) &
\xmark\ ($p=0.793$) &
\cmark\ ($p<0.001$) \\

GPT-4.1-mini &
\cmark\ ($p<0.001$) &
\xmark\ ($p=0.777$) &
\cmark\ ($p<0.001$) \\

GPT-5-mini &
\cmark\ ($p<0.001$) &
\xmark\ ($p=0.212$) &
\cmark\ ($p<0.001$) \\

Llama 3.3 Instruct (70B) &
\cmark\ ($p=0.001$) &
\xmark\ ($p=0.160$) &
\cmark\ ($p<0.001$) \\
\bottomrule
\end{tabular}
}
\end{table}

\finding{Exposing the implementation under test consistently reduces the fault detection effectiveness of LLM-generated test suites. Providing both the task description and the implementation decreases fault detection by 13.2\% on average compared to using the task description alone, while providing only the implementation results in an average decrease of 15.1\%.}

\subsection{RQ2: Prompting}

\begin{figure}[tb]
    \centering
    \includegraphics[width=1\linewidth]{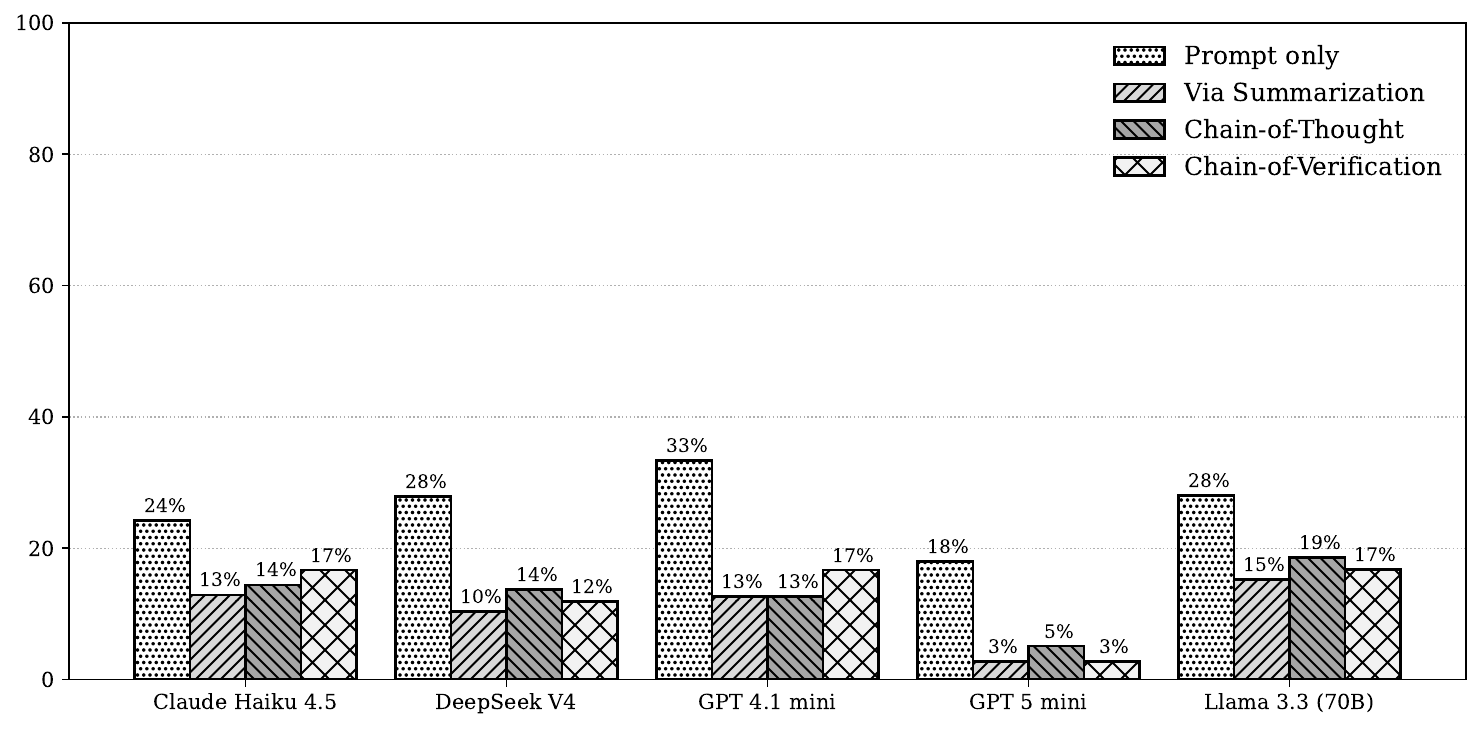}
    \caption{\textbf{RQ2: Fault Detection using Prompt Engineering Techniques.}}
    \label{fig:fd_rq2_technique}
\end{figure}

Figure \ref{fig:fd_rq2_technique} illustrates the fault detection obtain for each prompt engineering technique used. We observe that generating test suites from the task description alone is still more effective than any approach contains the code under test. More precisely, Prompt-only achieves higher fault detection than the summarization-based workflow across all evaluated models, with improvements of 11.4\%, 17.5\%, 20.7\%, 15.3\%, and 12.8\% for Claude Haiku 4.5, DeepSeek V4, GPT-4.1 mini, GPT-5 mini, and Llama 3.3 (70B), respectively. A similar trend is observed for Chain-of-Thought, where Prompt-only improves fault detection by 9.8\%, 14.1\%, 25.8\%, 12.9\%, and 9.5\%, respectively. Likewise, compared to Chain-of-Verification, Prompt-only achieves improvements of 7.6\%, 16.0\%, 16.7\%, 15.3\%, and 11.3\% across the same models.

\finding{Generating test suites directly from the task description is more effective than prompt engineering techniques, with fault detection decreasing by 15.5\% for summarization and 13.4\% for both Chain-of-Thought and Chain-of-Verification.}

\begin{table}[bt]
\centering
\caption{RQ2: Statistical significance between Prompt-only and the three prompting techniques.}
\label{tab:rq2_stats}

\resizebox{\columnwidth}{!}{%
\begin{tabular}{lccc}
\toprule
\textbf{Model} &
\textbf{Summ. vs. Prompt only} &
\textbf{CoT vs. Prompt only} &
\textbf{CoVe vs. Prompt only} \\
\midrule
Claude Haiku 4.5 &
\cmark\ ($p=0.0008$) &
\cmark\ ($p=0.0042$) &
\cmark\ ($p=0.0311$) \\

DeepSeek V4 &
\cmark\ ($p<0.0001$) &
\cmark\ ($p=0.0001$) &
\cmark\ ($p<0.0001$) \\

GPT-4.1-mini &
\cmark\ ($p<0.0001$) &
\cmark\ ($p<0.0001$) &
\cmark\ ($p=0.0001$) \\

GPT-5-mini &
\cmark\ ($p<0.0001$) &
\cmark\ ($p<0.0001$) &
\cmark\ ($p<0.0001$) \\

Llama 3.3 Instruct (70B) &
\cmark\ ($p=0.0001$) &
\cmark\ ($p=0.0042$) &
\cmark\ ($p=0.0005$) \\
\bottomrule
\end{tabular}
}
\end{table}

\subsection{RQ3: Workflows}

\begin{figure}[t]
    \centering
    \includegraphics[width=1\linewidth]{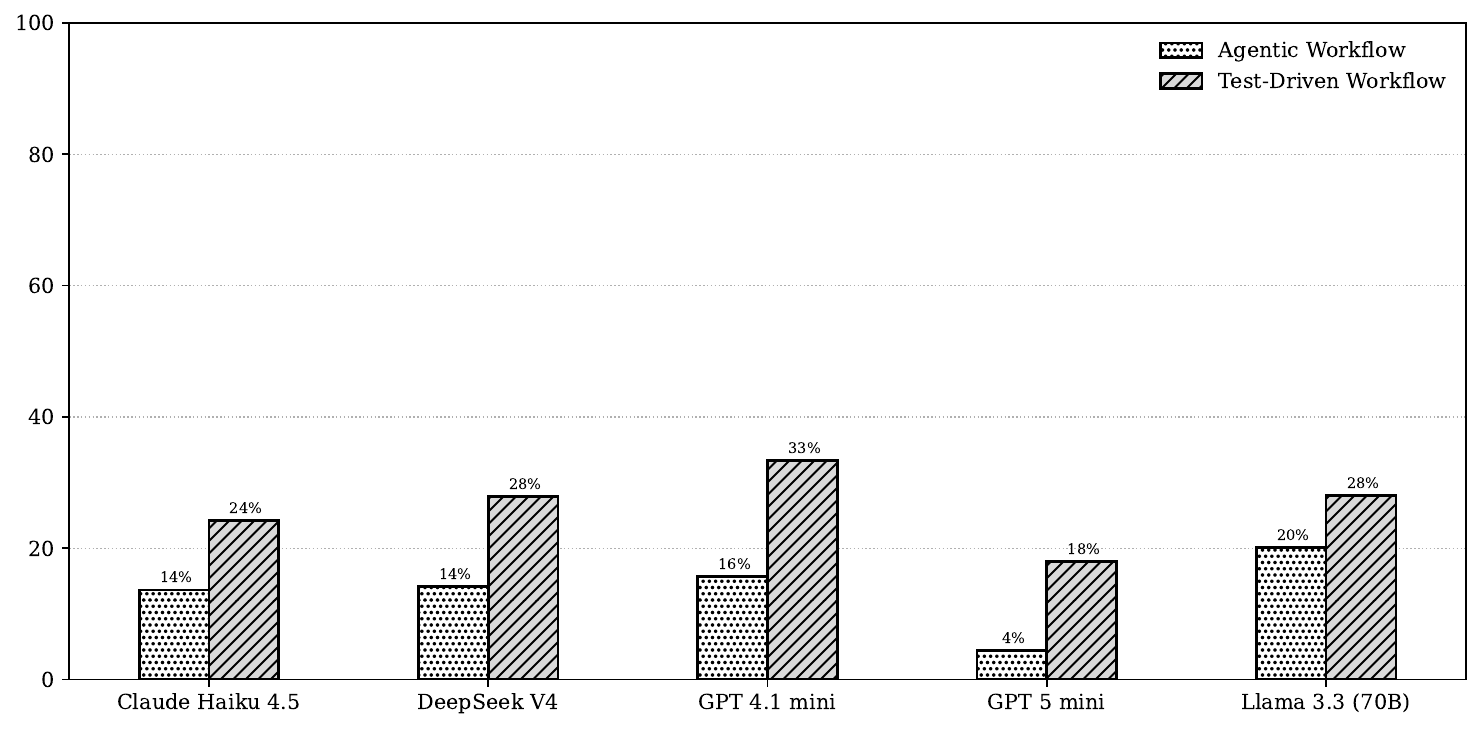}
    \caption{\textbf{Fault Detection under different workflows}} 
    \label{fig:fd_workflows}
\end{figure}

Figure~\ref{fig:fd_workflows} presents the fault detection effectiveness of the Agentic Workflow and the Test-Driven Workflow across all evaluated benchmarks. Consistent with the findings of RQ1, all evaluated models achieve higher fault detection rates under the Test-Driven Workflow than under the Agentic Workflow. Specifically, the Test-Driven Workflow improves fault detection by 10.6\% for Claude Haiku 4.5, 13.8\% for DeepSeek V4, 17.7\% for GPT-4.1-mini, 13.6\% for GPT-5-mini, and 7.9\% for Llama 3.3 (70B), relative to the Agentic Workflow. All differences are statistically significant as reported in Table \ref{tab:rq3_stats}

The observed trend is consistent across all evaluated models and does not appear to depend on whether reasoning capabilities are employed. For example, GPT-5-mini and DeepSeek V4 exhibit improvements of approximately 14\%, but only the former employs reasoning which we used. 

\finding{Test-Driven Workflow consistently generates test suites with higher fault detection effectiveness than the Agentic Workflow, detecting 11.7\% more faults on average.}

\begin{table}[t]
\centering
\caption{RQ3: Statistical significance of the comparison between the agentic workflow and the test-driven workflow}
\label{tab:rq3_stats}

\begin{tabular}{lc}
\toprule
\textbf{Model} &
\textbf{Agentic vs. Test-Driven} \\
\midrule
Claude Haiku 4.5 &
\cmark\ ($p=0.0019$) \\

DeepSeek V4 &
\cmark\ ($p=0.0001$) \\

GPT-4.1-mini &
\cmark\ ($p<0.0001$) \\

GPT-5-mini &
\cmark\ ($p<0.0001$) \\

Llama 3.3 Instruct (70B) &
\cmark\ ($p=0.0177$) \\
\bottomrule
\end{tabular}

\end{table}

\section{Controlling for Under-specified prompts, coverage, fault triggering and number of tests}

\subsection{Qualitative analysis}

 Our empirical analysis evaluate the effectiveness of the test driven workflow in terms of fault detection. However, this analysis does not explain \emph{why} certain strategies perform better than others. To better understand the observed differences, we conduct a diagnostic analysis using additional metrics that characterize the generated test suites.

Specifically, we investigate three complementary aspects. First, we measure the \emph{number of generated test cases}, as previous work has suggested that certain prompting strategies lead the LLM to produce more comprehensive test suites~\cite{konstantinou2026llmbasedtestgenerationtechniques}, which may influence fault detection effectiveness. Second, we measure \emph{statement coverage} to determine whether differences in code coverage explain the observed performance gap as shown in prior work~\cite{chekam2017empirical}. 

Finally, we evaluate \emph{fault triggering} to determine whether the generated test suites primarily validate the current implementation, a phenomenon previously observed in LLM-generated test oracles~\cite{konstantinou2024llmsgeneratetestoracles}.

\begin{table}[b]
\centering
\small
\caption{Total number of generated test cases per model and technique. The highest value in each row is shown in \textbf{bold}, while the second highest is \underline{underlined}.}
\label{tab:test_counts}

\resizebox{\columnwidth}{!}{%
\begin{tabular}{lrrrrrrr}
\toprule
\textbf{Model} &
\textbf{Summarization} &
\textbf{Prompt} &
\textbf{P + C} &
\textbf{Code} &
\textbf{CoT} &
\textbf{CoVe} &
\textbf{Agentic} \\
\midrule

Claude Haiku 4.5 &
4133 &
4374 &
4516 &
4488 &
\underline{5860} &
\textbf{7789} &
4727 \\

DeepSeek V4 Flash &
2111 &
2361 &
2489 &
2449 &
\underline{2912} &
\textbf{3262} &
2480 \\

GPT-4.1 Mini &
1221 &
1406 &
1398 &
1395 &
\underline{1727} &
\textbf{1953} &
1255 \\

GPT-5 Mini &
1971 &
1521 &
1951 &
1875 &
\underline{2126} &
\textbf{2701} &
2086 \\

Llama 3.3 70B Instruct &
2297 &
2295 &
2262 &
2244 &
\underline{2714} &
\textbf{3487} &
2351 \\

\bottomrule
\end{tabular}%
}
\end{table}

Table~\ref{tab:test_counts} reports the total number of tests generated by each model. Overall, generating tests directly from the task description (Prompt-only) does not produce the largest test suites. Instead, the prompting techniques CoVe and CoT consistently generate more tests across all models. Nevertheless, despite producing larger test suites, neither technique outperforms the Prompt-only approach on fault detection effectiveness. This observation is particularly evident for Claude Haiku 4.5, where CoVe generates almost double the number of tests, yet Prompt-only still achieves better fault detection. These findings suggest that the effectiveness of generating tests directly from the task description cannot be explained by the number of test cases.

Table~\ref{tab:line_coverage} reports the average statement (line) coverage achieved by the generated test suites for each model-configuration combination. Intuitively, test suites with higher coverage, exercise more execution paths which might lead to better fault detection. Although there is ongoing research on the most effective test adequacy criterion for fault detection~\cite{chekam2017empirical}, studying statement coverage and fault detection relationship may reveal whether coverage explains the observed differences. Overall, we observe no consistent relationship between statement coverage and fault detection. While generating tests directly from the task description achieves the highest statement coverage for DeepSeek V4, this pattern does not hold for the remaining models. Likewise, configurations with similar or even higher statement coverage do not necessarily achieve higher fault detection rates. Therefore, it is unlikely that the observed differences can be attributed to the portion of covered lines.

\begin{figure}[h]
    \centering
    \includegraphics[width=1\linewidth]{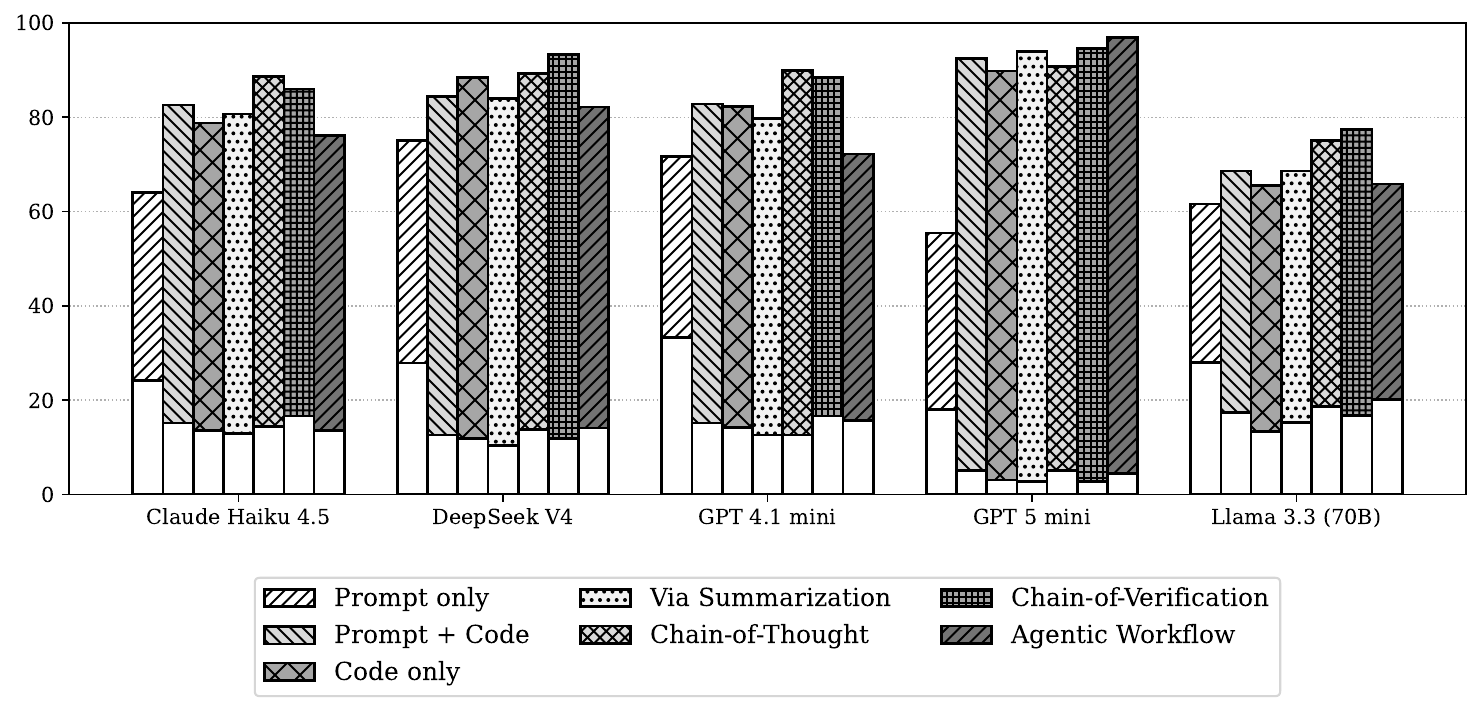}
    \caption{\textbf{Fault Detection and Fault Triggering}. Bottom values denote fault detection, while the stacked bars report the corresponding fault triggering rate.}
    \label{fig:fd_and_ft}
\end{figure}

Figure~\ref{fig:fd_and_ft} illustrates the fault detection rates reported in our results, together with their corresponding fault triggering rates. We observe that fault triggering and fault detection are not consistently correlated. For example, generating test suites directly from the task description achieves the highest fault detection rates across models, yet it often exhibits the lowest fault triggering rates. However, the agentic workflow achieves higher fault detection than generating tests from the implementation alone, but its fault triggering rates are also higher in all models except GPT-5 mini. 

\finding{Based on the qualitative analysis, we cannot attribute the observed differences in fault detection to the number of tests, statement coverage, or fault triggering. These results suggest that the effectiveness of the test suites is likely influenced by the quality of the generated assertions. While our study does not directly evaluate oracle correctness, these findings suggest that improving the quality of generated test oracles may be a promising direction for future work.}

\begin{table}[t]
\centering
\caption{Average line coverage across all models and test generation workflows. The highest value for each model is shown in \textbf{bold}, while the second highest is \underline{underlined}.}
\label{tab:line_coverage}

\resizebox{\columnwidth}{!}{%
\begin{tabular}{lccccccc}
\toprule
\textbf{Model} &
\textbf{Summarization} &
\textbf{Prompt} &
\textbf{P + C} &
\textbf{Code} &
\textbf{CoT} &
\textbf{CoVe} &
\textbf{Agentic} \\
\midrule
Claude Haiku 4.5 &
97.31\% &
97.26\% &
97.51\% &
94.69\% &
\textbf{97.90\%} &
\underline{97.81\%} &
97.37\% \\

DeepSeek V4 &
98.38\% &
\textbf{99.19\%} &
98.94\% &
95.56\% &
\underline{99.17\%} &
98.87\% &
98.78\% \\

GPT-4.1-mini &
97.85\% &
\underline{98.12\%} &
98.00\% &
95.13\% &
\textbf{98.20\%} &
97.91\% &
96.60\% \\

GPT-5-mini &
96.70\% &
96.63\% &
96.72\% &
82.84\% &
\underline{96.97\%} &
\textbf{97.71\%} &
95.22\% \\

Llama 3.3 Instruct (70B) &
95.91\% &
95.73\% &
95.26\% &
92.76\% &
\textbf{96.15\%} &
94.96\% &
\underline{96.04\%} \\
\bottomrule
\end{tabular}
}
\end{table}
\subsection{Fault detection using Under-specified prompts}

Our results indicate that generating tests from the task specification alone consistently outperforms workflows that expose the implementation under test. This observation raises a natural question: how does the fault detection effectiveness of LLM-generated test suites change when the task specification itself is under-specified?

Prior study explored the influence of under-specified prompts~\cite{akli2026promptunderspecificationimprovescode} when it comes to the correctness of generated code. The authors introduced three prompt variants that reduce the semantic information provided in the task specification. 

In this study, we adopt the same prompt variants to examine whether prompt underspecification similarly affects the fault detection effectiveness of LLM-generated test suites. 

We perform this analysis on the MBPP benchmark, for which Akli et al.~\cite{akli2026promptunderspecificationimprovescode} provide the three under-specified variants of each task description. Using these mutated prompts, we repeat the test generation experiments for all evaluated models. As in our primary evaluation, the LLM receives only the task description as input. However, in this experiment, the original prompt is replaced with its corresponding under-specified variant.

\begin{figure}[t]
    \centering
    \includegraphics[width=1\linewidth]{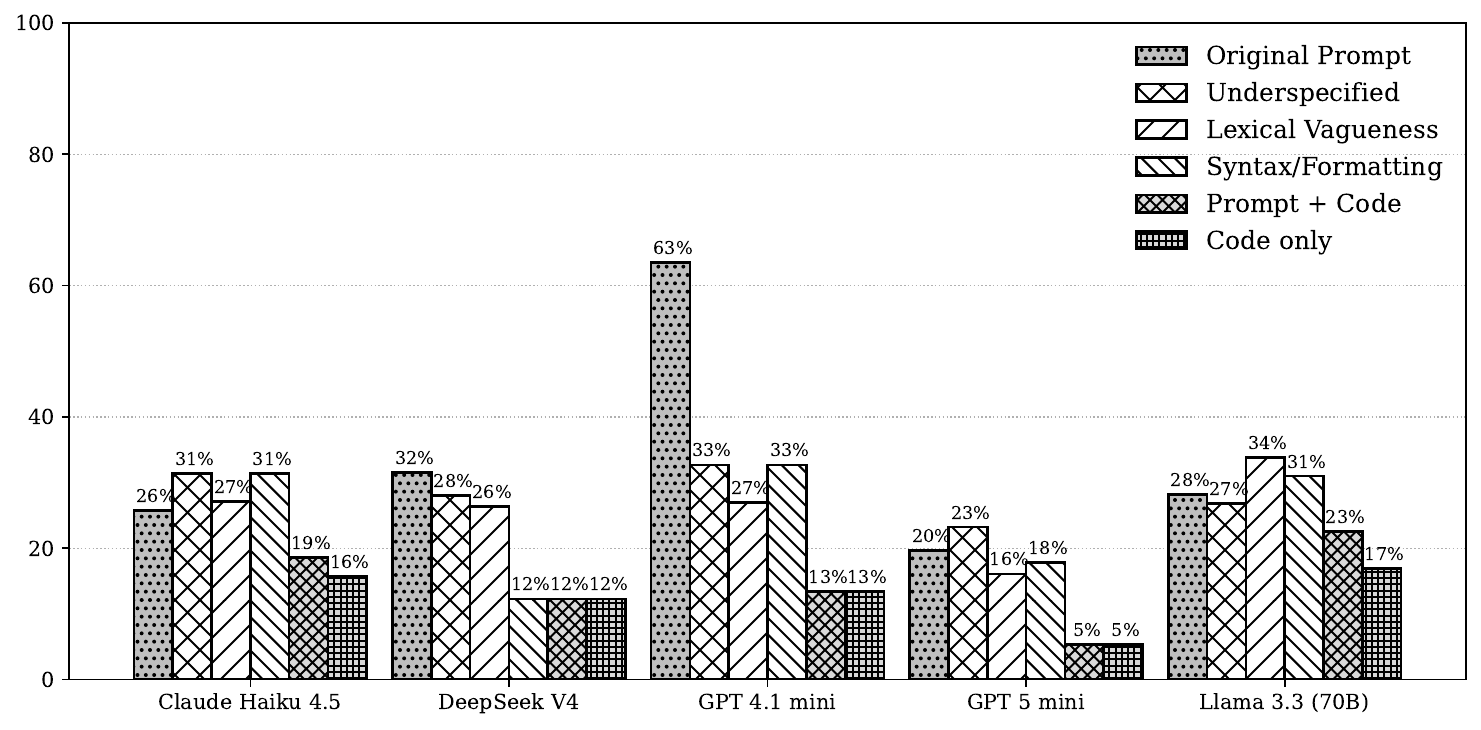}
    \caption{\textbf{Fault Detection on MBPP tasks using weak prompts}} 
    \label{fig:fd_weak_prompts}
\end{figure}

Figure~\ref{fig:fd_weak_prompts} presents the fault detection effectiveness on MBPP when test suites are generated from the original task description, from the implementation under test, and from the progressively under-specified task descriptions. Consistent with the findings of RQ1, test suites generated solely from the task description consistently outperform those generated from the implementation across all evaluated models. This trend persists even when the task description is deliberately weakened, suggesting that the semantic information retained in the under-specified prompts remains more beneficial for test generation than direct exposure to the implementation under test. Specifically, under-specified lexically vague, and modified syntax-formatting prompts had higher average fault detection over the prompt and code workflow by 14.0\%, 11.6\%, and 10.6\% respectively.

An additional observation is that, for some models, test suites generated from the under-specified prompts detected more faults than those generated from the original task descriptions. Although these differences are not sufficiently consistent to conclude that under-specified prompts are generally superior, they are nevertheless aligned with the findings of Akli et al.~\cite{akli2026promptunderspecificationimprovescode}. The authors show that task descriptions may inadvertently confuse the model instead of guiding it, and under-specified prompts allow the model to reason and infer the intended behavior.

\finding{Regardless of the quality of the task description, generating test suites directly from the task description yields higher fault detection effectiveness than generating tests from the implementation.}

\section{Related Work}

Overall, existing literature primarily evaluates the quality of LLM-generated tests using metrics such as coverage, compilation rate, or fault-detection capability.  While these studies establish the feasibility of using LLMs for test generation, they generally assume that the generated implementation is fixed and correct, and they do not investigate how the ordering of activities—generating code before tests versus generating tests before code—affects the ability of tests to reveal implementation defects. Our work addresses this gap by empirically studying the code-before-test workflow. Rather than focusing solely on test quality, we examine whether generating tests after code creation increases the likelihood that tests inherit the same assumptions, misunderstandings, or defects present in the implementation. Consequently, our study provides new insights into the risks of LLM-assisted testing within the development workflows.

\subsection{Empirical study on test generations}

Several studies investigated the capability of LLMs to generate tests \cite{konstantinou2026llmbasedtestgenerationtechniques, HuangZHDC26, Yuan0DW00L24}. Konstantinou et al. \cite{konstantinou2026llmbasedtestgenerationtechniques} investigated the performance of four representative execution-feedback (LLM-based) test generation methods (HITS~\cite{WangL0J24}, SymPrompt~\cite{10.1145/3643769}, TestSpark~\cite{sapozhnikov2024testspark} and Coverup~\cite{abs-2403-16218}) against a simple LLM-plain baseline. They showed that LLM-plain outperforms existing  methods when using newer LLMs. Yuan et al. \cite{Yuan0DW00L24} studied unit test generations with ChatGPT, finding that many of the tests lead to compilation errors and therefore proposed integrating compilation feedback to reduce them. Abdullin et al. \cite{AbdullinDP25} compared traditional test generation methods such as symbolic execution or SBST against LLM-based methods. They observed that LLM-based method could outperform more traditional ones in terms of mutation score. 

All the above studies aim at generating regression tests that assert the observed program behavior and therefore, by construction, cannot detect faults in the target implementations under test \cite{konstantinou2024llmsgeneratetestoracles,HuangZHDC26}. To detect faults, tests need to be equipped with oracles that assert the expected program behavior. To this end, TOGA \cite{10.1145/3510003.3510141} and TOGLL \cite{hossain2024togllcorrectstrongtest} use Evosuite to generate test prefixes, and leverage LLMs to generate the test oracles. They found that LLMs generate many wrong assertions. Konstantinou et al. \cite{konstantinou2024llmsgeneratetestoracles} showed that the inclusion of the faulty code in the LLM prompt negatively biasses the correctness of the generated assertions making them hard to detect faults. Going a step further, Huang et al. \cite{HuangZHDC26} showed that using the faulty code together with the task descriptions (or correct code) improves the fault detection ability of tests/assertions. 

Similarly to the above studies, we use LLMs to generate tests and oracles from the task descriptions, without including any faulty code in the candidate prompts to avoid potential bias. Differently though, rather than checking the impact of including a faulty implementation in the prompt, we examine the indirect impact of having a code generation during an earlier step within the agentic workflow. 

\subsection{LLMs for Code Generation}

A large body of work has investigated the use of LLMs for program synthesis. Many strategies and prompt engineering techniques have been proposed ~\cite{DBLP:conf/iclr/Sclar0TS24, DBLP:conf/iclr/SunSW24, DBLP:conf/emnlp/IsmithdeenKK25, DBLP:journals/tacl/MizrahiKMDSS24,DBLP:conf/emnlp/ZhuoZFDL024, DBLP:conf/emnlp/ChatterjeeRB024} with the intention to effectively support code generation. Recent studies have shown that LLM outputs may be influenced by specification ambiguity, prompting effects, and model biases \cite{larbi2025promptsgowrong, akli2026defective}. Research studies have also shown that simple perturbations such as syntactic transformations to the prompts~\cite{DBLP:conf/iclr/Sclar0TS24, DBLP:conf/emnlp/ZhuoZFDL024}, semantically preserving instruction rephrasing~\cite{DBLP:conf/iclr/SunSW24, DBLP:conf/emnlp/ChatterjeeRB024}, and other linguistic modifications~\cite{DBLP:conf/emnlp/IsmithdeenKK25,fagadau2024copilot,ma2025promptstability,DBLP:conf/acl/VoronovWR24}, can impact the correctness of the model's answers. Even when code appears correct according to generated tests, hidden behavioral deficiencies may remain undetected \cite{yang2024evaluationlargelanguagemodels}. These findings raise concerns about relying exclusively on LLM-generated tests as validation artifacts and motivate a closer examination of the interaction between code generation and test generation. 

\subsection{Test-First and Test-Driven Development}

The software engineering community has long advocated test-first development and test-driven development (TDD) as mechanisms for improving software quality and reducing defects. In traditional TDD, developers write tests before implementation, ensuring that requirements are explicitly captured and verified during development. The success of TDD is largely attributed to the independence between specification (tests) and implementation, which reduces confirmation bias and encourages requirement clarification \cite{beck2000extreme}.

The emergence of LLMs challenges this assumption. In many practical workflows, developers first generate code using an LLM and subsequently ask either the same model or another model to produce tests for that generated code. While this ``code-before-test'' workflow is convenient, it potentially introduces a strong dependence between implementation and generated tests. Because the tests are conditioned on the generated code, they may validate the implementation rather than the intended specification, creating a form of oracle bias. Despite extensive research on LLM-based test generation, relatively little work has studied the risks associated with generating tests after implementation as we do in this paper.

\section{Threats to Validity}

A primary threat stems from the non-deterministic nature of LLMs, which may produce different implementations and test suites across executions. To mitigate this threat, we use low-temperature settings whenever possible and repeat generation ten times when higher temperatures are required.

The generalizability of our findings may be limited by the choice of tasks, models, and experimental settings. The benchmark problems used in our study may not fully represent the diversity and complexity of real-world software engineering projects. Industrial systems often involve larger codebases, evolving requirements, domain-specific constraints, and extensive human oversight, which could affect the interaction between generated implementations and generated tests.  

Additionally, our results are based on a specific set of large language models and prompting strategies, designed to reflect realistic software development workflows. Alternative prompting strategies and workflows could lead to different behaviors and therefore different conclusions. For instance, test first generation, iterative refinement, agent-based verification, or workflows involving multiple independent models, may yield different outcomes and should not be assumed to exhibit the same limitations identified in this work.

\section{Conclusion}
Large language models are increasingly being integrated into software development workflows, where they are used not only to generate code but also to generate the tests intended to validate that code. While this workflow offers substantial productivity benefits, it raises a fundamental concern: does generating tests after code lead to tests that validate the implementation rather than the intended specification?

In this paper, we conducted an empirical study of the LLM-based code-before-test workflow and investigated its impact on the effectiveness of generated test suites. Our results indicate that tests generated after implementation are often strongly influenced by the code they observe, resulting in a tendency to reinforce implementation assumptions and overlook defects. Consequently, high test pass rates and code coverage do not necessarily imply that the generated implementation conforms to the original specification. Instead, such tests may provide a false sense of correctness by validating the same interpretation adopted during code generation.

The findings highlight a previously underexplored risk of LLM-assisted development: the loss of independence between implementation and verification. This issue becomes particularly problematic when task descriptions are ambiguous or underspecified, as both the generated code and the generated tests may converge to the same incorrect interpretation. In such cases, the testing process fails to serve its primary purpose of identifying behavioral deviations from the intended requirements.

Our study contributes empirical evidence that the ordering of code and test generation matters in LLM-driven software engineering workflows. The results suggest that practitioners should exercise caution when relying exclusively on post-hoc LLM-generated tests as correctness oracles and should prefer workflows that preserve a degree of separation between specification, implementation, and validation.

Future work can explore techniques to mitigate this risk, including specification-first prompting, independent test generation models, adversarial test generation, and human-in-the-loop validation strategies. More broadly, we believe that understanding the interactions between multiple LLM-generated artifacts will be essential for building trustworthy AI-assisted software engineering processes. As LLMs continue to assume greater responsibility in development workflows, ensuring that testing remains an independent and effective mechanism for quality assurance will be critical in realizing their potential.

\bibliographystyle{IEEEtran}
\bibliography{main.bib}

\end{document}